# Quantum dynamics simulation of the advection-diffusion equation

Hirad Alipanah,[1] Feng Zhang,[2,3] Yong-Xin Yao,[2,3] Richard Thompson,[4] Nam Nguyen,[5] Junyu Liu,[6] Peyman Givi,[1,7] Brian J. McDermott,[8] and Juan José Mendoza-Arenas[1,9,*]

[1]*Department of Mechanical Engineering and Materials Science, University of Pittsburgh, Pittsburgh, Pennsylvania 15261, USA*
[2]*Ames National Laboratory, U.S. Department of Energy, Ames, Iowa 50011, USA*
[3]*Department of Physics and Astronomy, Iowa State University, Ames, Iowa 50011, USA*
[4]*Integrated Vehicle Systems, Applied Mathematics, Boeing Research & Technology, Huntsville, Alabama 35824, USA*
[5]*Integrated Vehicle Systems, Applied Mathematics, Boeing Research & Technology, Huntington Beach, California 92647, USA*
[6]*Department of Computer Science, University of Pittsburgh, Pittsburgh, Pennsylvania 15261, USA*
[7]*Department of Petroleum Engineering, University of Pittsburgh, Pittsburgh, Pennsylvania 15261, USA*
[8]*Naval Nuclear Laboratory, Schenectady, New York 12301, USA*
[9]*Department of Physics and Astronomy, University of Pittsburgh, Pittsburgh, Pennsylvania 15261, USA*

The advection-diffusion equation is simulated via several quantum algorithms. Three formulations are considered: (1) Trotterization, (2) variational quantum time evolution (VarQTE), and (3) adaptive variational quantum dynamics simulation (AVQDS). These schemes were originally developed for the Hamiltonian simulation of many-body quantum systems. The finite-difference discretized operator of the transport equation is formulated as a Hamiltonian and solved without the need for ancillary qubits. Computations are conducted on a quantum simulator (IBM Qiskit Aer) and a superconducting quantum hardware (IBM Fez). The former emulates the latter without the noise. The actual hardware implementation experiences significant noise. The results of the quantum simulator are compared with data from direct numerical simulation (DNS) with infidelities of the order $10^{-5}$. In the quantum simulator, Trotterization is observed to have the lowest infidelity and is suitable for fault-tolerant computation. The AVQDS algorithm requires the lowest gate count and circuit depth. The VarQTE algorithm is the next best in terms of gate counts, but the number of its optimization variables is directly proportional to the number of qubits. Due to current hardware limitations, Trotterization cannot be implemented, as it has an overwhelmingly large number of operations. Meanwhile, AVQDS and VarQTE can be executed at the hardware level. These algorithms present a new paradigm for computational transport phenomena on quantum computers.

## I. INTRODUCTION

Quantum computing (QC) is now recognized as a promising tool for numerical simulations of transport phenomena in science and engineering [1–11]. Traditional methods for solving the partial differential equations (PDEs) that govern such processes often require significant runtime and memory resources [12], especially when dealing with high-dimensional and high-resolution systems [13]. Quantum computing is expected to enable simulations that can overcome these challenges by leveraging quantum superposition, entanglement, and the exponential scalability of the Hilbert space with the number of qubits. These attributes of quantum computing, in turn, enable efficient encoding and manipulation of high-order tensor representations of the solution field, potentially yielding quantum speedups over classical methods [14–19]. Significant progress has been made in the development of algorithms with potential quantum speedups [20,21]. However, quantum advantages can only be achieved on ideal, fault-tolerant quantum computers. Until such hardware becomes available, noisy intermediate-scale quantum (NISQ) hardware [22–25] provides an interim alternative to benchmark the conceptual frameworks of larger-scale methods.

In recent years, variational quantum algorithms (VQAs) [26–28] have shown promise in simulating transport phenomena on NISQ devices [16,29–53]. These algorithms are hybrid quantum-classical schemes in which the transport equations are transformed into cost functions to be minimized. The quantum processor encodes and evaluates trial solutions by choosing the rotation angles (i.e., variational parameters) in an ansatz, while a classical optimizer iteratively adjusts the circuit parameters to minimize the residual. The VQA approach leverages the ability of variational quantum circuits to efficiently explore large-state spaces using a relatively small number of qubits, making them well suited for computationally intense classical problems. However, as the number of qubits increases, the optimization landscape can become exponentially flat, leading to vanishing gradients. This phenomenon, known as *barren plateau*, presents challenges in designing optimal quantum circuits [54–57] and will add challenges to the optimization process. Another challenge of the VQA approach is to obtain state tomography. Here, at each time step, the quantum state must be measured and reconstructed for use in the subsequent step. This process poses a bottleneck for solving differential equations using QC [58].

[*]Contact author: jum151@pitt.edu

The objective of this work is to explore the potentials of QC for numerical simulation of the advection-diffusion equation [48,53,59–65]. This equation serves as a convenient testbed for assessing the applicability of quantum algorithms to transport phenomena, including fluid mechanics, heat and mass transfer, combustion, and many others. Three QC methods are considered: Trotterization [66–68], variational quantum time evolution (VarQTE) [53,69–75], and adaptive variational quantum dynamics simulation (AVQDS) [76]. These schemes were originally developed for simulating the Hamiltonian dynamics of many-body quantum systems. Trotterization [77] involves decomposing the quantum evolution operator into smaller, implementable operators by approximating the exponential of the sum of noncommuting terms in the Hamiltonian as a product of exponentials. This method, commonly used in digital quantum simulations, enables the approximation of nonunitary operations using unitary gates. VarQTE [69] and AVQDS [76] leverage variational principles to simulate temporal evolution at both real and imaginary times. The distinction between these methods lies in the structure of the ansatz employed for optimization. VarQTE uses a fixed ansatz chosen at the onset. AVQDS employs an adaptive ansatz in which the operators change dynamically as time evolves. These methods facilitate simulations on current NISQ devices by reducing the circuit depth and the number of gates to a level much lower than that of Trotterization. The ansätze are also implemented on the IBM Fez quantum hardware to assess their practical viability.

## II. FORMULATION

Transport of a conserved scalar is considered under the influence of convection and diffusion. This scalar is denoted by $C(x,t)$, where $0 \leqslant x \leqslant L$ and $t \geqslant 0$ denote the physical space and time, respectively. Convection is induced via a constant velocity $U$, and the diffusion is assumed Fickian with a constant diffusion coefficient $\Gamma$. The space is normalized by $L$, and the time is normalized by $\frac{L}{U}$. In this setting, the scalar transport is governed by

$$\frac{\partial C}{\partial t} + \frac{\partial C}{\partial x} = \frac{1}{Pe}\frac{\partial^2 C}{\partial x^2}, \tag{1}$$

where the dimensionless Péclet number ($Pe = \frac{LU}{\Gamma}$) provides a measure of advection to diffusion. For numerical computations, the spatial derivatives are discretized via a second-order central finite difference scheme

$$\begin{aligned}\frac{\partial C(x_i)}{\partial t} &+ \frac{C(x_{i+1}) - C(x_{i-1})}{2\Delta x}\\ &= \frac{1}{Pe}\frac{C(x_{i+1}) - 2C(x_i) + C(x_{i-1})}{\Delta x^2},\end{aligned} \tag{2}$$

where $x_i$ $(i = 0, 1, \ldots 2^N - 1)$ denotes the grid points. The function $C$ is evaluated on an exponentially large number of grid points. The finite difference scheme of Eq. (2) allows $C$ to be naturally encoded into a quantum register of just $N$ qubits. The wavefunction $|C\rangle$ of the set of qubits is defined by having $C(x_i)$ as the $i$th element of its vector representation. With the assumption of periodic boundary conditions in $x$, the wave-function transport is given by

$$\frac{\partial |C\rangle}{\partial t} = \hat{A}|C\rangle, \tag{3}$$

with the non-Hermitian "Hamiltonian-like" operator $\hat{A}$ defined as

$$\hat{A} = \frac{1}{Pe\Delta x}\begin{pmatrix} b & c & 0 & 0 & 0 & \dots & 0 & d\\ d & b & c & 0 & 0 & \dots & 0 & 0\\ 0 & d & b & c & 0 & \dots & 0 & 0\\ 0 & 0 & d & b & c & \dots & 0 & 0\\ \vdots & \vdots & \vdots & \vdots & \vdots & & \vdots & \vdots\\ c & 0 & 0 & 0 & 0 & \dots & d & b \end{pmatrix},$$

$$b = -\frac{2}{\Delta x}, \quad c = \frac{1}{\Delta x} - \frac{Pe}{2}, \quad d = \frac{1}{\Delta x} + \frac{Pe}{2}. \tag{4}$$

Equation (3) describes a nonunitary evolution and does not preserve the norm of $|C\rangle$. The nonunitary time evolution operator cannot be directly implemented with quantum gates (i.e., unitary operators). By defining the imaginary time as $\beta = it$, Eq. (3) is expressed in the form of the Schrödinger equation

$$\frac{\partial |C\rangle}{\partial \beta} = -i\hat{A}|C\rangle, \tag{5}$$

with Hamiltonian $\hat{H} = -\hat{A}$. Equation (5) is the subject of QC. This Hamiltonian can be separated into its Hermitian and an anti-Hermitian components. The advection and the diffusion terms in Eq. (1) create the anti-Hermitian and the Hermitian components, respectively. The three quantum algorithms used to solve Eq. (5) are described below in order.

### A. Quantum imaginary time evolution via Trotterization

Many problems in quantum mechanics are expressed as equations of the form (3), commonly referred to as Schrödinger-like equations in imaginary time. Examples include: calculating the thermal state of a quantum system at a specific temperature [78], finding the ground state of a Hamiltonian in the long-time limit [79,80], and simulating the dynamics and steady state of open quantum systems using a Lindblad master equation [78,81]. Considering these problems in the context of many-body quantum systems poses significant challenges due to strong correlations. The quantum imaginary time evolution (QITE) algorithm [66,67,82,83] was introduced to harness QC to simulate equations of the form (3). The implementation of the QITE algorithm begins by expressing the solution to the discretized form of Eq. (1) as

$$|C(t)\rangle = e^{i\hat{H}\beta}|C(0)\rangle = e^{-\hat{H}t}|C(0)\rangle. \tag{6}$$

This time evolution cannot be implemented directly on a quantum hardware, since quantum gates invoke unitary operators. Moreover, Eq. (6) does not preserve the norm of the quantum state $|C(t)\rangle$. By breaking down the full evolution into a number of small time steps, each individual step can be represented by a unitary evolution together with a normalization

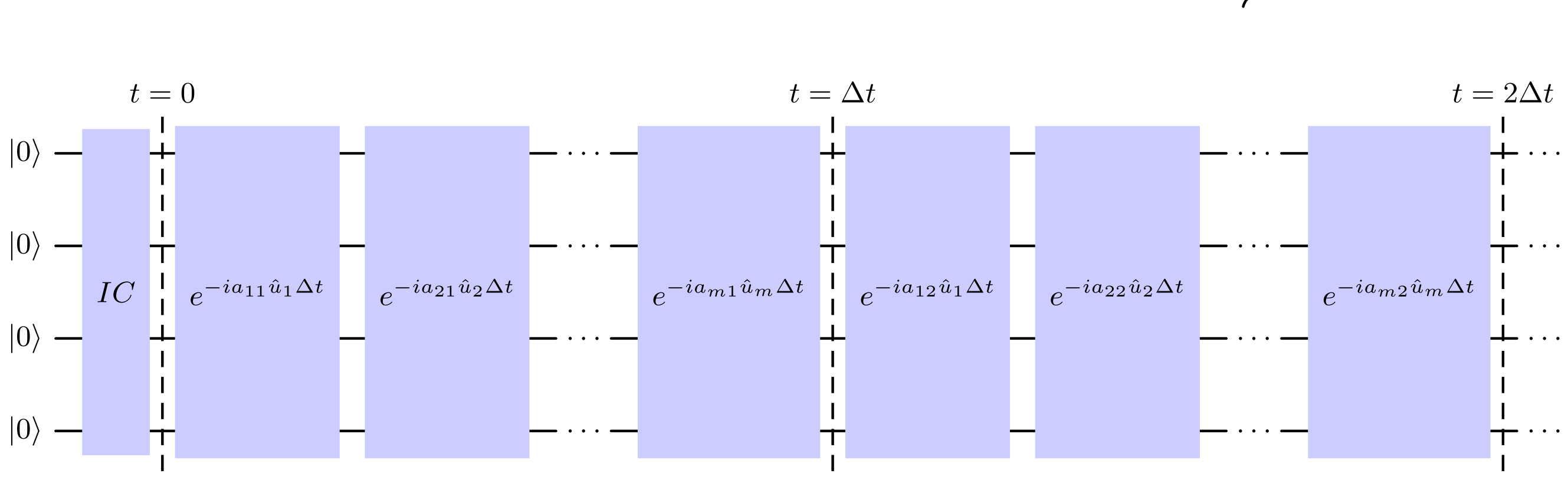

FIG. 1. Trotterized imaginary time evolution performed on 4 qubits using $k$ time steps. The circuit begins with the initial condition $IC$. At each step, the coefficients $a_{jk}$ are calculated by Eq. (8b). Afterwards, the unitaries $e^{-ia_{jk}\hat{u}_j\Delta t}$ are added to the circuit. Each dashed vertical line indicates the evolution time.

factor

$$
\begin{aligned}
|C(t+\Delta t)\rangle &= \frac{e^{-\hat{H}\Delta t}|C(t)\rangle}{\|e^{-\hat{H}\Delta t}|C(t)\rangle\|} \\
&= \frac{e^{-\hat{H}\Delta t}|C(t)\rangle}{\sqrt{\langle C(t)|e^{-(\hat{H}+\hat{H}^\dagger)\Delta t}|C(t)\rangle}} \\
&\approx e^{-i\hat{U}\Delta t}|C(t)\rangle.
\end{aligned} \tag{7}
$$

In practice, this approximation is performed term by term to approximate the action of each finite-range Pauli term in the Hamiltonian by a full-range unitary operator. The Pauli term decomposition of this Hamiltonian is described in Sec. III. In Eq. (7), the nonunitary time evolution operator is approximated by a unitary operator $e^{-i\hat{U}\Delta t}$. Therefore, the problem translates into finding the suitable $\hat{U}$. This can be done by decomposing $\hat{U}$ into a linear combination of $m$ Pauli strings as $\hat{U} = \sum_{j=1}^{m} a_j \hat{u}_j$. The coefficients $a_j$ are obtained by solving the equation $S\vec{a} = \vec{b}$ classically, where

$$
S_{jl} = \langle C(t)|\hat{u}_j^\dagger \hat{u}_l|C(t)\rangle, \tag{8a}
$$

$$
\begin{aligned}
b_j &= \frac{-i\langle C(t)|\hat{u}_j^\dagger \hat{H}|C(t)\rangle}{\sqrt{\langle C(t)|e^{-(\hat{H}+\hat{H}^\dagger)\Delta t}|C(t)\rangle}} \\
&\approx \frac{-i\langle C(t)|\hat{u}_j^\dagger \hat{H}|C(t)\rangle}{\sqrt{1-\Delta t\langle C(t)|(\hat{H}+\hat{H}^\dagger)|C(t)\rangle}}.
\end{aligned} \tag{8b}
$$

The accuracy of approximating the nonunitary evolution operator in Eq. (7) depends on the number of qubits utilized to encode the unitary operator. When the Pauli terms of $\hat{H}$ act on $p$ neighboring qubits, domains with $D > p$ qubits are required to encode the unitary approximation of the nonunitary evolution operator. Initial applications of the QITE algorithm to many-body quantum systems demonstrate rapid convergence to exact results as $D$ increases [66]. After determining the coefficients $a_j$, the time evolution of the entire system is decomposed into a sequence of evolution operations corresponding to each Pauli string. The decomposition is performed using a first-order Trotter expansion

$$
e^{-i\hat{U}_k\Delta t} \approx \prod_{j=1}^{m} e^{-ia_{jk}\hat{u}_j\Delta t} + \mathcal{O}(\Delta t^2). \tag{9}
$$

Here, $u_j$ denotes the $j$th Pauli string, $a_{jk}$ is its associated coefficient at time $t = k\Delta t$, and $\hat{U}_k = \sum_{j=1}^{m} a_{jk}\hat{u}_j$. Therefore, the full evolution after $k$th time step is

$$
\begin{aligned}
|C(t)\rangle &= (e^{-\hat{H}\Delta t})^{\frac{t}{\Delta t}}|C(0)\rangle \\
&\approx \left(\prod_{k=1}^{\frac{t}{\Delta t}} \frac{1}{c_k} \prod_{j=1}^{m} e^{-ia_{jk}\hat{u}_j\Delta t}\right)|C(0)\rangle,
\end{aligned} \tag{10}
$$

with $c_k$ denoting the norm of $|C(k\Delta t)\rangle$. The overall procedure is outlined in Fig. 1. At each time step, the unitaries associated with each block are applied to the quantum state at the previous step. Therefore, the circuit depth and the number of gates scale linearly with the total number of time steps.

Due to its simplicity, Trotterization has been one of the preferred approaches to simulate the time evolution of correlated quantum systems on both classical [79,80,84] and quantum devices [85–88]. However, the circuit depth in this method depends on the complexity of the problem. For local Hamiltonians, some $a_{jk}$ values become negligible. On the other hand, nonlocal Hamiltonians make use of all $a_{jk}$ values. Hence, the circuit depth for complex Hamiltonians can grow rapidly with time, making this method feasible only on fault-tolerant hardware.

### B. Variational quantum time evolution

With this formulation, the time evolution of the set of variational parameters $\vec{\theta}$ is of primary interest. These parameters characterize a fixed ansatz employed for the solution of Eq. (3). Several variational approaches have been proposed [69]. Here, McLachlan's scheme [89] is employed, in which the *McLachlan's distance* between the left-hand side and the right-hand side of the Schrödinger equation is minimized.

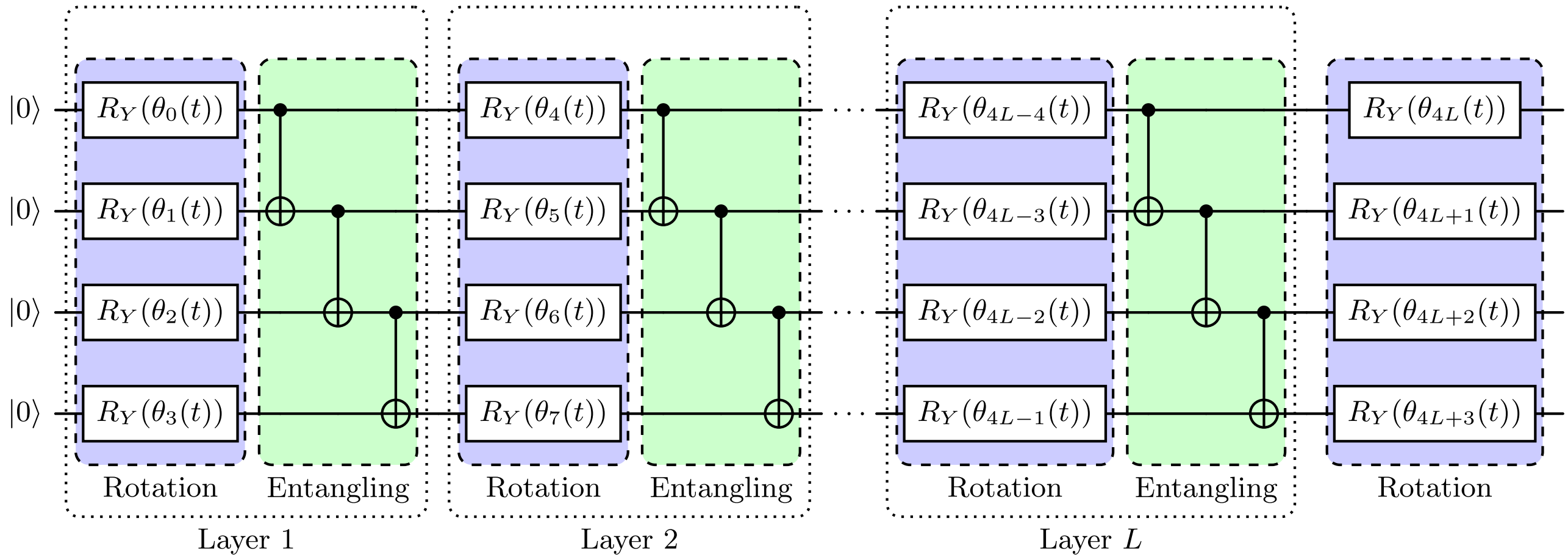

FIG. 2. VarQTE ansatz with 4 qubits and $L$ layers. In each layer, $R_Y$ gates create a rotation component that changes the real-valued amplitudes, and the entangling component creates local correlations between the qubits. The structure is repeated $L$ times and is terminated with a final rotation. The parameters $\theta_j(t)$ are used for minimizing the McLachlan distance. The evolved state at any time $t$ is obtained by measuring the quantum state after applying the ansatz with parameters $\theta_j(t)$ to the initial zero state.

Thus, this approach is based on finding

$$\min_{\dot{\vec{\theta}}} \left\| \frac{\partial |C(\vec{\theta})\rangle}{\partial t} + (\hat{H} - \langle \hat{H} \rangle_t)|C(\vec{\theta})\rangle \right\|, \tag{11}$$

where $\langle \hat{H} \rangle_t = \langle C(\vec{\theta}(t))|\hat{H}|C(\vec{\theta}(t))\rangle$ denotes the expected value of $\hat{H}$ at time $t$. The problem of finding the optimal $\theta(t)$ for Hermitian and anti-Hermitian Hamiltonians has already been considered in Ref. [69]. Here, the solutions of both cases are combined to obtain the evolution under an arbitrary operator. The Hamiltonian $\hat{H}$ is decomposed into its Hermitian and anti-Hermitian components $\hat{H} = \hat{H}_1 + i\hat{H}_2$, where $H_1$ and $H_2$ (both Hermitian) correspond to the diffusion and advection terms, respectively. With this decomposition, the problem translates into minimizing

$$D = \left\| \frac{\partial |C(\vec{\theta})\rangle}{\partial t} + \tilde{H}_1|C(\vec{\theta})\rangle + i\tilde{H}_2|C(\vec{\theta})\rangle \right\|, \tag{12}$$

with respect to $\dot{\vec{\theta}}$, where $\tilde{H}_j = \hat{H}_j - \langle \hat{H}_j \rangle_t$ for $j \in \{1, 2\}$. The term $\langle \hat{H}_1 \rangle_t$ shifts the Hamiltonian $\hat{H}_1$ to enforce normalization, and $\langle \hat{H}_2 \rangle_t$ shifts the Hamiltonian $\hat{H}_2$ to reduce the rapid global phase oscillation. The dynamics induced by the Hermitian and anti-Hermitian components are solved separately through imaginary time evolution and real time evolution, respectively. Thus, the problem is reduced to solving the linear system $A\dot{\vec{\theta}} = R$ [69] with coefficients

$$A_{jk} = \mathrm{Re}\left[ \frac{\partial \langle C(\theta(t))|}{\partial \theta_j(t)} \frac{\partial |C(\theta(t))\rangle}{\partial \theta_k(t)} + \frac{\partial \langle C(\theta(t))|}{\partial \theta_j(t)} \times |C(\theta(t))\rangle \langle C(\theta(t))| \frac{\partial |C(\theta(t))\rangle}{\partial \theta_k(t)} \right], \tag{13a}$$

$$R_j = \mathrm{Im}\left[ \frac{\partial \langle C(\theta(t))|}{\partial \theta_j(t)} \hat{H}|C(\theta(t))\rangle - \langle \hat{H} \rangle_t \frac{\partial \langle C(\theta(t))|}{\partial \theta_j(t)} |C(\theta(t))\rangle \right]. \tag{13b}$$

Equations (13a) and (13b) are time dependent. Therefore, an iterative method is required to determine $\vec{\theta}$. The evolution from $\vec{\theta}(t)$ to $\vec{\theta}(t + \Delta t)$ consists of the following steps: (1) extract the coefficients of $A$ and $R$ by running Hadamard test-type circuits [90], (2) solve the linear systems of equations $A\dot{\vec{\theta}} = R$ classically, (3) advance $\theta$ from $t$ to $t + \Delta t$ by the numerical solution of $A\dot{\vec{\theta}} = R$. Using the forward Euler scheme for the last step

$$\vec{\theta}(t + \Delta t) = \vec{\theta}(t) + \dot{\vec{\theta}}\Delta t = \vec{\theta}(t) + A^{-1}R\Delta t.$$

The ansatz for VarQTE is shown in Fig. 2. This ansatz consists of $L$ layers, each containing rotation ($R_Y$) and entangling ($CX$) gates. The rotation operations are performed about the $y$ axis to keep the outcomes real valued. After the $L$ layers, another rotation component is applied to introduce additional $N$ degrees of freedom. Therefore, the total number of parameters (rotation angles) to optimize in an $N$ qubit ansatz is $N(L + 1)$. The circuit depth of each layer is 3, resulting in a total depth of $3L + 1$. The initial values are specified so that the variational quantum state matches the actual initial conditions $C(x, 0)$. This is implemented through amplitude embedding and sequential least squares programming (SLSQP) minimization [72]. The VarQTE method limits dealing with vanishing gradients to only the initial state, as the subsequent steps do not optimize any cost functions.

### C. Adaptive variational quantum dynamics simulation

The accuracy of VarQTE is often constrained by the fixed variational ansatz due to its limited degrees of freedom. The AVQDS algorithm [76,91] overcomes this limitation by leveraging McLachlan's distance $D$, which quantifies the discrepancy between the dynamical trajectories of variational and exact simulations. When $D$ exceeds a preset threshold $D_{\max}$, new parametrized unitaries are appended to the ansatz to reduce $D$. These unitaries are generated from a predefined pool of Pauli terms, $\{\hat{I}, \hat{X}, \hat{Y}, \hat{Z}\}^{\otimes N}$, and are selected based on their effectiveness in minimizing $D$. Multiple unitaries can be attached at each time step until $D < D_{\max}$. This way, the variational state adopts the form

$$|C(\vec{\theta})\rangle = \prod_{j=0}^{N_\theta - 1} e^{-i\theta_j \hat{\mathcal{A}}_j} |C(\vec{\theta}(t_0))\rangle, \tag{14}$$

where $N_\theta$ denotes the number of operators selected from the pool, and $\hat{\mathcal{A}}_j$ denote the corresponding Pauli strings chosen from the pool. The number $N_\theta$ depends on the Hamiltonian structure and does not necessarily depend on the number of qubits. To improve the efficiency, the pool is restricted to rotations applying real-valued operators to the state, which requires Pauli terms with an odd number of $\hat{Y}$ operators. This constraint approximately halves the pool size, significantly accelerating classical implementation of the algorithm. The procedure, outlined in Fig. 3, begins with standard VarQTE time evolution. When $D$ exceeds $D_{\max}$ at a time step $t_1$, the distance is computed for all pool terms, which are then scored based on their distance reduction. The highest-scoring term, $\mathcal{A}_1$, is selected, and its exponential is added to the circuit. This adaptive process repeats until the final time $T$.

## III. SIMULATIONS

### A. Pauli string decomposition

To implement time-evolution algorithms, the Hamiltonian needs to be decomposed into a linear combination of Pauli strings: the identity, a left-shift operator, and a right-shift operator. The latter two operators shift the vector representation of a quantum state by $\Delta x$ either to the left or to the right. The shift operators are necessary for the implementation of algorithms based on finite-difference schemes. For instance, using $C_+(x)$ as the notation for the left-shifted function and $C_-(x)$ for the right-shifted function, the second-order central difference operator is

$$\begin{aligned}\frac{\partial^2 C}{\partial x^2} &\approx \frac{C(x_{i+1}) - 2C(x_i) + C(x_{i-1})}{\Delta x^2}\\ &= \frac{C_+(x_i) - 2C(x_i) + C_-(x_i)}{\Delta x^2}.\end{aligned} \tag{15}$$

Using $N$ qubits, the left-shift operator is expressed as the $2^N \times 2^N$ matrix:

$$\hat{T}_N = \begin{pmatrix} 0 & 1 & 0 & 0 & 0 & \dots & 0 & 0\\ 0 & 0 & 1 & 0 & 0 & \dots & 0 & 0\\ 0 & 0 & 0 & 1 & 0 & \dots & 0 & 0\\ 0 & 0 & 0 & 0 & 1 & \dots & 0 & 0\\ \vdots & \vdots & \vdots & \vdots & \vdots & \ddots & \vdots & \vdots\\ 1 & 0 & 0 & 0 & 0 & \dots & 0 & 0 \end{pmatrix}. \tag{16}$$

The right-shift operator can be considered as a left-shift applied on the reverse-ordered quantum state (i.e., a vector with elements being the same as those of the original vector but ordered in reverse). Therefore, the right-shift operator is $\hat{T}_N^\dagger$. Higher-order finite difference algorithms can be implemented using powers of $\hat{T}_N$ and $\hat{T}_N^\dagger$. With these operators, the

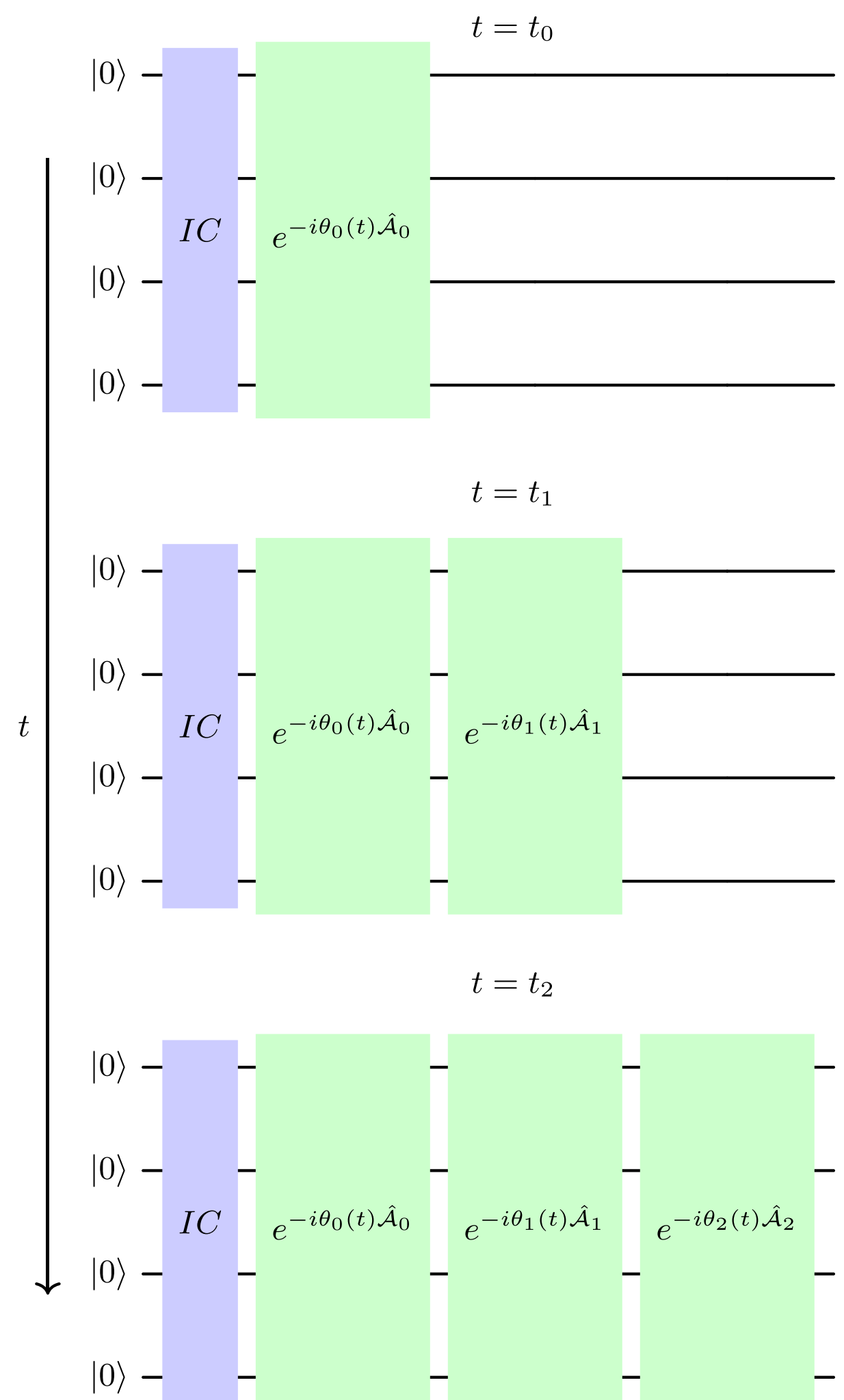

FIG. 3. AVQDS ansatz with 4 qubits and initial condition $IC$. At every time step, $D$ is calculated by Eq. (12) and is compared with $D_{\max}$. For the first instance $t = t_1$ when $D \geqslant D_{\max}$, the next term from the pool of operators is chosen and is attached to the ansatz. The ansatz continues to calculate the parameters until the next time for update at $t = t_2$. This process continues until the final time $T$. The evolved state at any time is obtained by measuring the quantum state after applying the ansatz with parameters $\theta_j(t)$ to the initial zero state.

Hamiltonian of Eq. (1) is

$$\begin{aligned}\hat{H} = \frac{1}{Pe}\bigg(&\frac{2}{\Delta x^2}\hat{I}^{\otimes N} - \left(\frac{1}{\Delta x^2} - \frac{Pe}{2\Delta x}\right)\hat{T}_N\\ &- \left(\frac{1}{\Delta x^2} + \frac{Pe}{2\Delta x}\right)\hat{T}_N^\dagger\bigg).\end{aligned} \tag{17}$$

Now, $\hat{T}_N$ is expressed in terms of Pauli strings. The creation ($\hat{a}^\dagger$) and annihilation ($\hat{a}$) operators, defined as

$$\hat{a} = \begin{pmatrix} 0 & 1\\ 0 & 0 \end{pmatrix} = \frac{1}{2}(\hat{X} + i\hat{Y}), \quad \hat{a}^\dagger = \begin{pmatrix} 0 & 0\\ 1 & 0 \end{pmatrix} = \frac{1}{2}(\hat{X} - i\hat{Y}), \tag{18}$$

can be used for this purpose. The left-shift operator $\hat{T}_N$ can be rewritten as

$$\hat{T}_N = \begin{pmatrix} \hat{a} & & & \\ & \hat{a} & & \\ & & \ddots & \\ & & & \hat{a} \end{pmatrix} + \begin{pmatrix} 0 & \hat{a}^\dagger & & \\ & 0 & \hat{a}^\dagger & \\ & & \ddots & \ddots \\ \hat{a}^\dagger & & & 0 \end{pmatrix}. \tag{19}$$

The first term has blocks of annihilation operators on the main diagonal. Therefore, this term is $\hat{I}^{\otimes N-1} \otimes \hat{a}$. The second term consists of blocks of creation operators with a pattern resembling that of a left-shift operator. Thus,

$$\hat{T}_N = \hat{I}^{\otimes N-1} \otimes \hat{a} + \hat{T}_{N-1} \otimes \hat{a}^\dagger. \tag{20}$$

This is a recursive expression for $\hat{T}_N$. With $\hat{T}_1 = \hat{X}$,

$$\begin{aligned} \hat{T}_N &= \hat{I}^{\otimes N-1} \otimes \hat{a} + \hat{I}^{\otimes N-2} \otimes \hat{a} \otimes \hat{a}^\dagger \\ &\quad + \hat{I}^{\otimes N-3} \otimes \hat{a} \otimes \hat{a}^\dagger \otimes \hat{a}^\dagger + \cdots \\ &\quad + \hat{I} \otimes \hat{a} \otimes (\hat{a}^\dagger)^{\otimes N-2} + \hat{X} \otimes (\hat{a}^\dagger)^{\otimes N-1} \\ &= \hat{I}^{\otimes N-1} \otimes \hat{a} + \sum_{j=1}^{N-2} \hat{I}^{\otimes N-1-j} \otimes \hat{a} \otimes (\hat{a}^\dagger)^{\otimes j} \\ &\quad + \hat{X} \otimes (\hat{a}^\dagger)^{\otimes N-1}. \end{aligned} \tag{21}$$

Similarly, for $\hat{T}_N^\dagger$,

$$\hat{T}_N^\dagger = \hat{I}^{\otimes N-1} \otimes \hat{a}^\dagger + \sum_{j=1}^{N-2} \hat{I}^{\otimes N-1-j} \otimes \hat{a}^\dagger \otimes \hat{a}^{\otimes j} + \hat{X} \otimes \hat{a}^{\otimes N-1}. \tag{22}$$

Thus, the final form of the Hamiltonian is

$$\begin{aligned} \hat{H} = \frac{1}{Pe}\Bigg[ & \frac{2}{\Delta x^2}\hat{I}^{\otimes N} - \left(\frac{1}{\Delta x^2} - \frac{Pe}{2\Delta x}\right)\left(\hat{I}^{\otimes N-1} \otimes \hat{a} + \sum_{j=1}^{N-2} \hat{I}^{\otimes N-1-j} \otimes \hat{a} \otimes (\hat{a}^\dagger)^{\otimes j} + \hat{X} \otimes (\hat{a}^\dagger)^{\otimes N-1}\right) \\ & - \left(\frac{1}{\Delta x^2} + \frac{Pe}{2\Delta x}\right)\left(\hat{I}^{\otimes N-1} \otimes \hat{a}^\dagger + \sum_{j=1}^{N-2} \hat{I}^{\otimes N-1-j} \otimes \hat{a}^\dagger \otimes \hat{a}^{\otimes j} + \hat{X} \otimes \hat{a}^{\otimes N-1}\right)\Bigg]. \end{aligned} \tag{23}$$

The total number of Pauli terms for each component of the Hamiltonian is presented in Table I. This table shows that the number of Pauli strings scales exponentially with $N$. In one-dimensional (1D) systems, this might not lead to a significant quantum advantage. However, in higher dimensions, the Hamiltonian gets sparser. For example, in two-dimensional (2D) cases, only twice this number of Pauli terms is used to describe the Hamiltonian (for the differentiation operators in the two dimensions), whereas the size of the Hamiltonian is squared. The decomposition of Eq. (23) breaks the Hamiltonian down to its Pauli basis. This Pauli representation allows implementation of the algorithm on quantum computers.

### B. Simulation results

The transport of the scalar $C(x, t)$ is simulated from the initial trapezoidal profile as shown in Fig. 4(a). The Péclet number $Pe = 32$, with $N = 4$ qubits and $\Delta t = 0.002$ to perform stable calculations. The initial condition is encoded using an amplitude embedding map, implemented via the SciPy package in Python.

TABLE I. Number of Pauli terms for 1D advection-diffusion Hamiltonian. The total number of terms in this Hamiltonian is $2^N + 2^{N-1} - 1$. This number is approximately the square root of $4^N$, the total, number of all the Pauli strings with $N$ qubits.

| Term | Number of Pauli strings |
|---|---|
| $\hat{I}^{\otimes N}$ | 1 |
| $\hat{I}^{\otimes N-1} \otimes \hat{a}$ or $\hat{I}^{\otimes N-1} \otimes \hat{a}^\dagger$ | 2 |
| $\hat{I}^{\otimes N-1-j} \otimes \hat{a} \otimes (\hat{a}^\dagger)^{\otimes j}$ or $\hat{I}^{\otimes N-1-j} \otimes \hat{a}^\dagger \otimes \hat{a}^{\otimes j}$ | $2^{j+1}$ |
| $\hat{X} \otimes (\hat{a}^\dagger)^{\otimes N-1}$ or $\hat{X} \otimes \hat{a}^{\otimes N-1}$ | $2^{N-1}$ |
| Sum | $2^N + 2^{N-1} - 1$ |

Simulations are conducted over a full resident period ($0 \leqslant t \leqslant 1$). The quantum state $|C(t)\rangle$ is calculated in the IBM Qiskit Aer noiseless simulator with full connectivity, using the three algorithms and $10^6$ shots (the maximum number allowed). This simulator uses the same basis gates of the IBM Fez digital quantum hardware. The algorithms are transpiled into native gates ($SX$, $X$, $RZ$, and $CZ$).

In Fig. 4(a), the temporal evolution of the scalar field is shown through Trotterization. The DNS results are shown for the final time only ($t = 1$). The agreement with Trotterization is excellent at all times. The results obtained via the VarQTE and AVQDS algorithms are indistinguishable from those of DNS and Trotterization by naked eyes, and thus not shown. The fidelity of the simulations is defined as $f(t) = |\langle C(t)|\psi\rangle|^2$, obtained from the normalized DNS result (encoded in the amplitudes of a state $|\psi\rangle$) and the simulated states $|C(t)\rangle$. Figure 4(b) shows the infidelity $1 - f(t)$ for the three algorithms. The Trotterization algorithm has almost a constant infidelity throughout the time evolution. It also shows the lowest infidelity among the algorithms at long times.

### C. Resource estimation

The native gate count of the quantum circuits for Trotterization (Fig. 1), VarQTE (Fig. 2), and AVQDS (Fig. 3) is given in Table II for simulations with $N = 4$ qubits, assuming a linear chain-like layout of the qubits. For VarQTE, $L = 10$ layers are employed. The numbers reported in Table II exclude the auxiliary Hadamard-test measurement circuits used to estimate the coefficients. The structure of these circuits depends on the instantaneous ansatz and the operator pool. The details on the number of circuits required for performing Hadamard-test measurements are provided in Appendix C. Implementing Trotterization requires an extremely large number of gates. The evolution operators on the right-hand side of Eq. (10) are 4-qubit unitary operators. The decomposition of these unitaries into the native gates of the Fez quantum computer results in a quantum circuit of significant depth. The total number of native gates of the circuit far exceeds the capabilities of current NISQ hardware, which is limited to approximately 100 entangling gates [92]. As a result, despite its effectiveness for simulation of correlated quantum systems, Trotterization is not suitable for solving PDEs until fault-tolerant quantum devices become available.

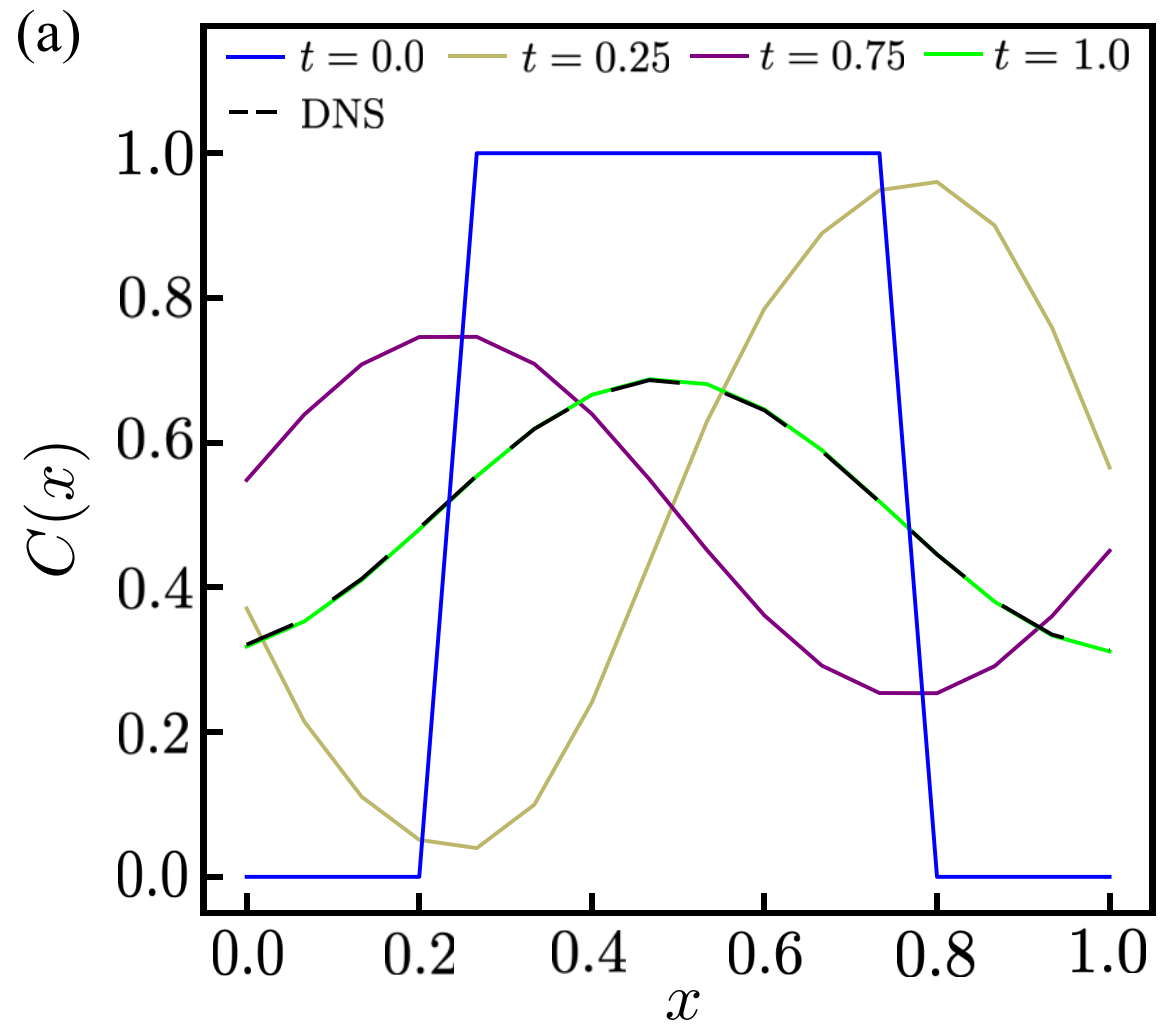

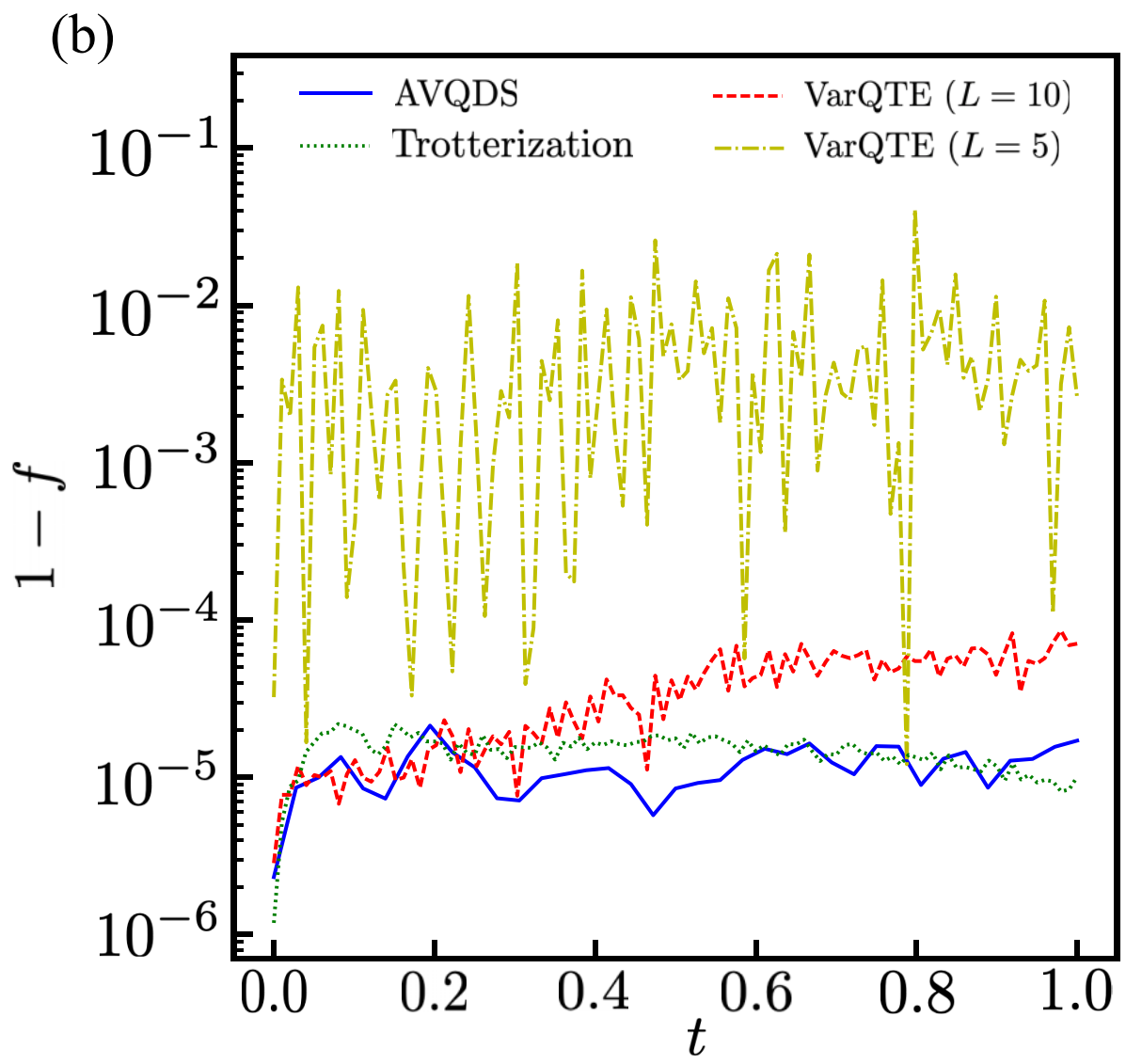

FIG. 4. (a) Evolution of the 1D advection-diffusion equation using Trotterization of imaginary time evolution. (b) Infidelity $(1-f)$ of the Trotterization, VarQTE with ansatz consists of 5 and 10 repetitions of the rotation-entangling layer, and AVQDS using an operator pool of Pauli terms consisting of odd number of $Y$ gates. The evolutions of the variational methods look almost the same as those of panel (a).

TABLE II. Circuit depth and operator count for the total simulations with $N = 4$ qubits, using the Trotterization, VarQTE, and AVQDS methods. The operators listed here are the native gates of the IBM Fez quantum computer.

| Gate | Trotterization | VarQTE ($L = 10$) | AVQDS |
|---|---|---|---|
| $X$ | 317 | 0 | 6 |
| $\sqrt{X}$ | 53 646 | 108 | 79 |
| $RZ$ | 48 460 | 109 | 67 |
| $CZ$ | 20 213 | 30 | 40 |
| Total | 122 636 | 247 | 192 |
| Depth | 76 021 | 90 | 129 |

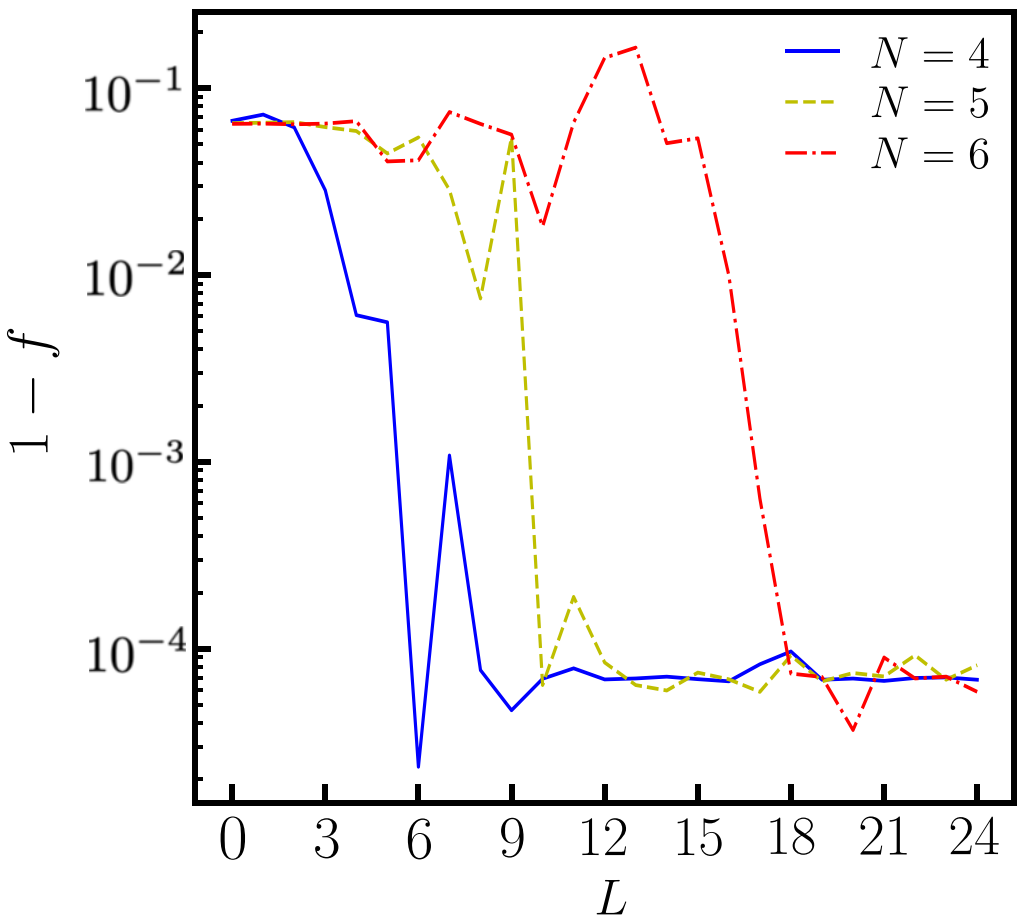

FIG. 5. The infidelity $(1-f)$ of VarQTE algorithm for $\Delta t = 0.002$, at the final evolution time $t = 1$, as a function of the number of layers $L$ for different number of qubits $N$.

The VarQTE algorithm requires significantly fewer gates than Trotterization. As shown in Fig. 4(b), the infidelity of VarQTE increases with time. However, as demonstrated in Fig. 5, the infidelity decreases significantly as the number of layers $L$ increases [72]. For $N = 4$ qubits, the infidelity is high with $L \leqslant 5$, and the simulations fail to capture the time evolution accurately. With $L = 10$ layers, the infidelity decreases to a value of order $\mathcal{O}(10^{-5})$ at final times. As the number of qubits increases (with a fixed time step $\Delta t$), more layers are required to achieve the same infidelity.

It should be noted that the AVQDS gate counts are also higher than the number of unitaries in the ansatz for the variational methods. This is the case because each unitary in the form of $e^{-i\theta\mathcal{A}}$, where $\mathcal{A}$ is a Pauli string, is transpiled into multiple native gates that can span multiple depth layers. It is possible to further restrict the choice of Pauli strings according to the geometry of the qubit layout. For instance, Pauli strings like $X_1Y_3$ can be removed from the operator pool in the AVQDS method since qubits 1 and 3 are not directly connected in the linear chain-like layout. Interestingly, numerical tests show that the final transpiled circuit assembled using this restricted pool has very similar total gate and depth counts to that used by the original pool containing all Pauli strings with odd number of Pauli-$Y'$s. This suggests that the final circuit captures the intrinsic complexity of the problem that transcends the specific choices of the operator pool.

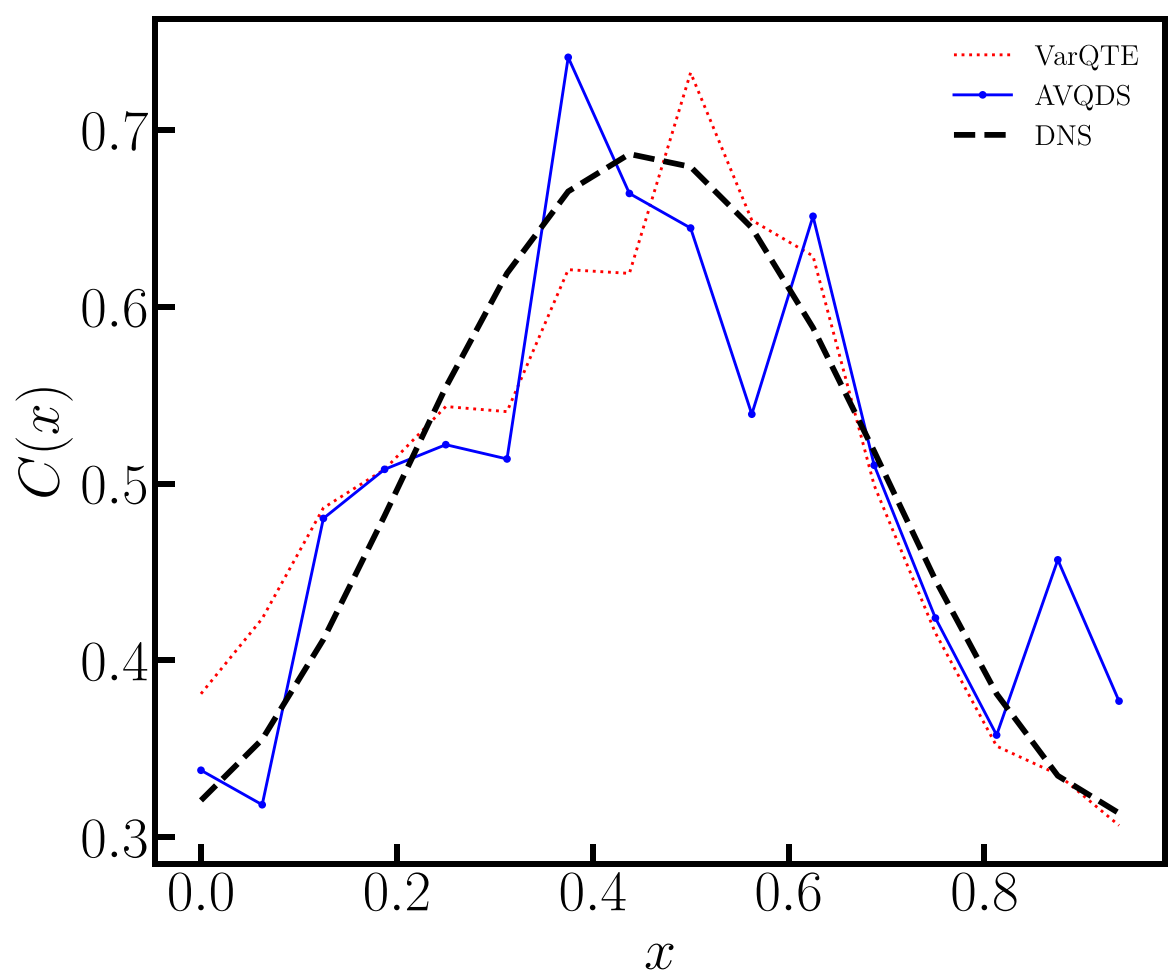

FIG. 6. Implementation of VarQTE and AVQDS ansätze for the final time of the simulation using the final parameters on IBM Fez. The simulations used 16 384 shots.

Similarly to VarQTE, the infidelity of the AVQDS algorithm increases slowly over time, reaching $\mathcal{O}(10^{-5})$ at the final time (see Fig. 4). However, AVQDS requires fewer gates than VarQTE for the same accuracy. Also, AVQDS is the fastest method that allows simulation with $N = 8$ qubits. Thus, AVQDS is rated as the best method to simulate the advection-diffusion equation.

### D. Implementation on quantum hardware

Variational simulation with $N = 4$ qubits with entangling gates is manageable on current NISQ devices. To assess the variational simulations' performance, the ansatz is implemented using the native gates on the target hardware. This is a valid check for the full variational simulation process, as the circuits required to perform Hadamard-type tests heavily depend on the ansatz itself. Therefore, in the quantum hardware experiments, the parameters for the final time step are obtained through noiseless simulation. Afterwards, using those parameters, both the VarQTE and AVQDS ansätze are implemented on the IBM Fez quantum computer. The IBM Fez is a 156-qubit quantum computer with a Heron r2 quantum processor. This quantum computer has a median longitudinal relaxation time (T1) of 131.07 μs and a median phase coherence time (T2) of 97.06 μs. The full ansatz for the VarQTE and AVQDS method are provided in Appendix B. The scalar profile at the final time with 16 384 shots is presented in Fig. 6. This is the maximum number of shots that IBM Fez uses. The scalar $C(x,t)$ only contains positive real values. Therefore, measurement in Pauli $Z$ basis is enough to obtain the full solution of the scalar [93].

The profiles generated by the two variational algorithms show the same trends as in DNS but with significant errors due to noise. The presence of such noise levels is widely and notoriously recognized in QC [1]. Error suppression schemes, namely, dynamical decoupling and Pauli twirling, were tested. The results of these schemes did not have a significant effect on the quality of the solution and were omitted from the presented results. The development of error correction schemes to eliminate or substantially reduce this error is the subject of significant current research [90,94–96].

### E. AVQDS on 2D systems

To demonstrate the generality and capability to more complex systems, simulations are extended to two-dimensional $(x - y)$ convection-diffusion. The evolution of the scalar $C(x, y, t)$ is considered on 4 qubits associated with each dimension (total of 8 qubits) implying a $16 \times 16$ discretization of the unit cell and a Péclet number $Pe = 132$. An $L$-shaped profile is imposed for the initial condition $C(x, y, 0)$. Only the AVQDS method is used as other methods requiring memory and processing power that exceeded available resources. Details on the vector representation of the 2D advection-diffusion equation are provided in Appendix A. Similar to the 1D case, the operator pool is selected by only selecting Pauli strings that contain an odd number of $\hat{Y}$ operators, so that $C(x, y, t)$ remains real during evolution. For a system with 8 qubits, the inclusion of all such Pauli strings results in a large pool with a total number of 32 640 operators. This will greatly affect the efficiency of the operator selection process for the adaptive construction of the ansatz. Thus, an additional constraint is imposed to limit the maximal length of the Pauli strings defined as *weight* $(w)$ in the pool. For instance, an operator pool with $w = 2$ consists of all operators $\hat{Y}_i$, $\hat{X}_i\hat{Y}_j$, and $\hat{Y}_i\hat{Z}_j$ with $1 \leqslant i, j \leqslant 8$, and $i \neq j$. In practice, the pools are constructed with $w = 2$, 3, and 4 containing 120, 848, and 3648 operators, respectively.

Figure 7 shows a few snapshots of the field from the simulation with the $w = 3$ operator pool. The results show the combined influence of convection and diffusion. An ansatz expressive enough to describe the dynamical evolution of this system cannot be assembled using the pool with $w = 2$. The results using pools with $w = 3$ and $w = 4$ yield excellent agreement with DNS, as shown by the infidelity results in Fig. 8. Simulations with higher $Pe$ values and/or extension to 3D are trivial, but require a higher number of qubits.

## IV. CONCLUSION

Three quantum algorithms, Trotterization, VarQTE, and AVQDS, are employed for the quantum simulation of the classical advection-diffusion equation. The finite-difference discretized form of this equation is cast in the form of a Hamiltonian and is decomposed into Pauli strings [97]. The implementation of the Hamiltonian is made possible by constructing an appropriate ansatz suitable for computations on NISQ machines. The AVQDS is shown to have the lowest gate counts among the three methods. The VarQTE method requires a relatively low depth, but the number of its parameters increases linearly with the number of qubits. The Trotterization is an all-quantum algorithm, but it is not currently possible to use it for simulations on quantum hardware. The other two algorithms can be employed, although they lead to significant errors due to the noisy nature of the existing

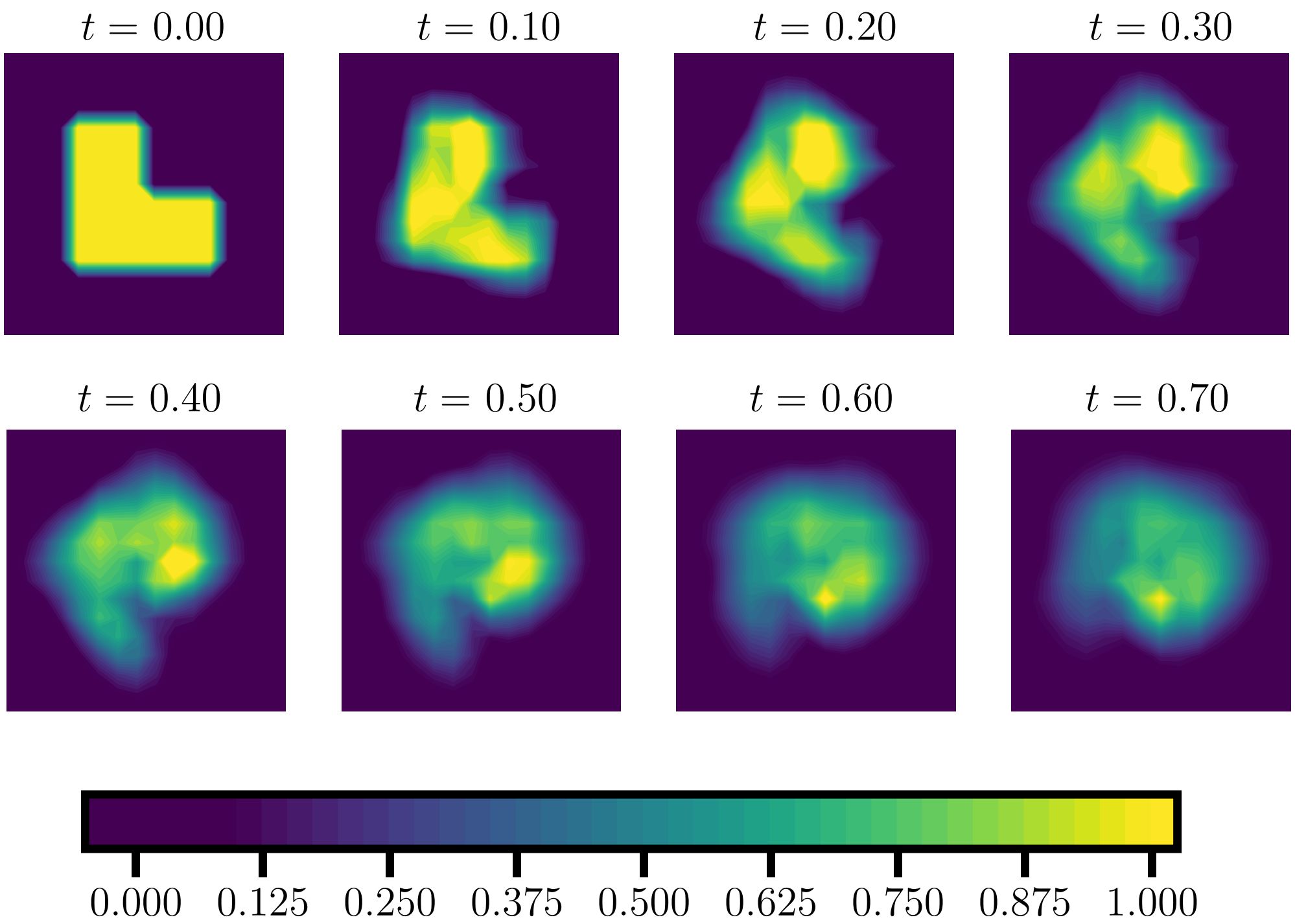

FIG. 7. Contour plots showing the evolution of the 2D scalar field $C(x, y, t)$ using 8 qubits (i.e., $16 \times 16$ grid) and $Pe = 132$ simulated using AVQDS. The DNS simulation with the same number of qubits resembles the same evolution (and not shown). The initial condition is depicted as an "L"-shaped function. The evolution shows this profile to rotate and diffuse.

hardware. Near-future quantum devices will allow calculations with much lower error [98].

This work provides a measure of the current capabilities of QC for simulating transport phenomena. The findings here also open up new avenues for future research, where potential quantum speedup can be harnessed to tackle problems currently beyond the reach of classical methods. Several paths for future research are suggested. One promising direction is the incorporation of nonlinearities into quantum circuits, by evaluating different strategies such as linearization methods [99,100] and nonlinear processing units [101]. This could facilitate the study of complex systems governed by nonlinear PDEs, such as the Navier-Stokes and/or the reaction-diffusion equations. Another avenue involves evaluating the solution of PDEs on alternative quantum computing platforms with digital or analog computing, including trapped ions and neutral atoms [102–105]. The determination of the most suitable algorithms for each platform remains an open question [8]. Lastly, enhancing algorithms by integrating tensor networks could improve computational efficiency and scalability [35,101,106–109], further extending the applicability of these methods.

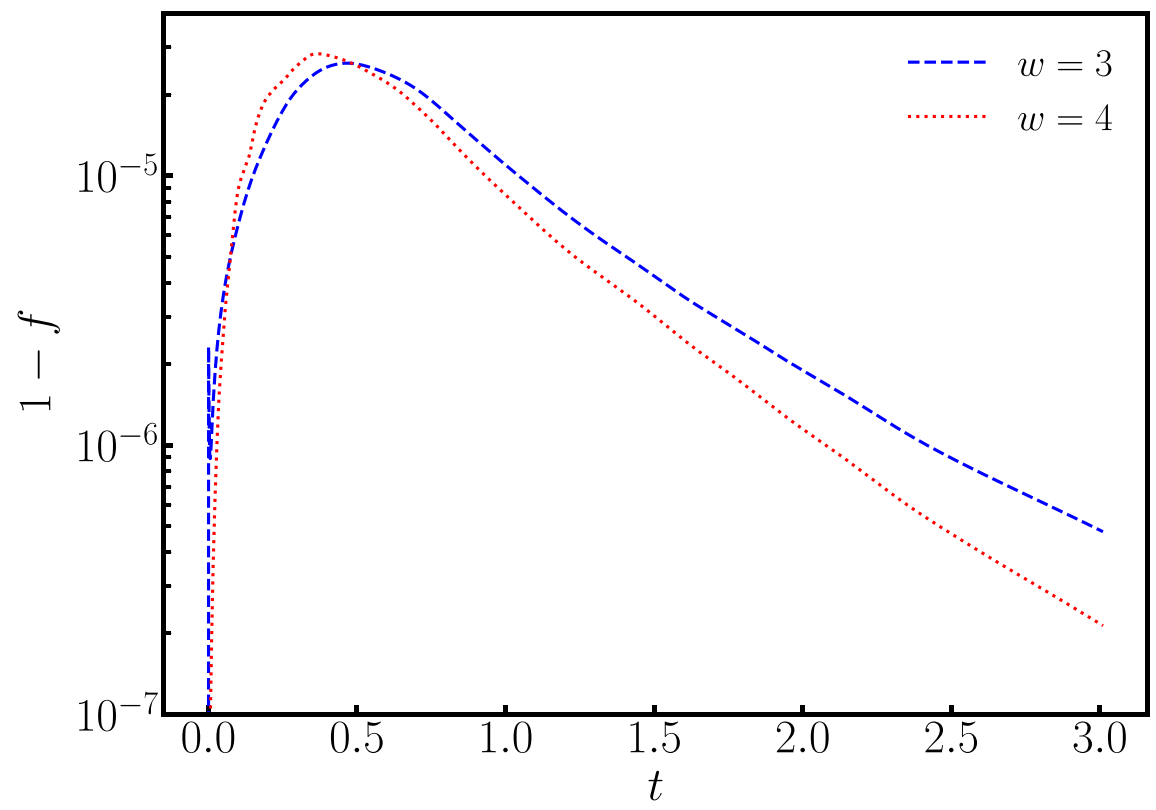

FIG. 8. The infidelity of the 2D wavefunction as a function of time during AVQDS with the weight of the operator pool $w = 3$ and $w = 4$. The low infidelity of $w = 3$ system indicates that this ansatz can be efficiently implemented using three-site gates on a quantum computer, which can further be decomposed into one and two-site native gates of the quantum hardware.

In the context of computational efficiency, current quantum algorithms are clearly not yet comparable to their classical counterparts. The expected improvements of quantum algorithms and quantum hardware will be crucial in expanding the applicability of quantum computing to complex problems.

## ACKNOWLEDGMENTS

The authors acknowledge support from the U.S. Air Force Office of Scientific Research (AFOSR) under Grant No. FA9550-23-1-0014. This research was supported in part by the University of Pittsburgh Center for Research Computing, RRID:SCR_022735, through the resources provided. Specifically, this work used the H2P cluster, which is supported by NSF Award No. OAC-2117681. J.L. was also supported in part by the Department of Computer Science at the University of Pittsburgh. The authors acknowledge the use of IBM quantum resources of the Air Force Research Laboratory. This work has been co-authored by a contractor of the U.S. Government under Contract No. DOE89233018CNR000004.

Accordingly, the U.S. Government retains a nonexclusive, royalty-free license to publish or reproduce the published form of this contribution, or allow others to do so, for U.S. Government purposes. The work by F.Z. and Y.Y. was supported by the U.S. Department of Energy (DOE), Office of Science, Basic Energy Sciences, Materials Science and Engineering Division, including the grant of computer time at the National Energy Research Scientific Computing Center (NERSC) in Berkeley, California. This part of the research was performed at the Ames National Laboratory, which is operated for the U.S. DOE by Iowa State University under Contract No. DE-AC02-07CH11358. The authors acknowledge discussions with B. Özgüler.

## DATA AVAILABILITY

The data that support the findings of this article are openly available [110]. The generalized AVQDS code as a quantum PDE solver can be accessed at [111], which is released as a submodule in the CyQC package [112]. The other algorithms have been implemented using the Qiskit algorithms library [113].

## APPENDIX A: 2D ADVECTION-DIFFUSION EQUATION

The conserved scalar in a 2D space is denoted by $C(x, y, t)$, where $-\frac{L_x}{2} \leqslant x \leqslant \frac{L_x}{2}$, $-\frac{L_y}{2} \leqslant y \leqslant \frac{L_y}{2}$, and $t \geqslant 0$ denote the physical space in two dimensions and time, respectively. Advection is through a velocity field $(U_x,\ U_y)$ and diffusion is assumed to be Fickian with a constant diffusion coefficient $\Gamma$. The 2D advection-diffusion equation is defined as

$$\frac{\partial C}{\partial t} + \frac{\partial U_x C}{\partial x} + \frac{\partial U_y C}{\partial y} = \Gamma\left(\frac{\partial^2 C}{\partial x^2} + \frac{\partial^2 C}{\partial y^2}\right). \tag{A1}$$

The velocity field yields a rotation $U_x = -\frac{y}{\sqrt{x^2+y^2}}$ and $U_y = \frac{x}{\sqrt{x^2+y^2}}$ so that the curl of the velocity field is constant. The discretized form of the transport via the central second-order scheme yields

$$\begin{aligned}\frac{\partial C(x_i, y_j)}{\partial t} = {} & \frac{y_j}{2\Delta x}\left(\frac{C(x_{i+1}, y_j)}{\sqrt{x_{i+1}^2 + y_j^2}} - \frac{C(x_{i-1}, y_j)}{\sqrt{x_{i-1}^2 + y_j^2}}\right) \\ & - \frac{x_i}{2\Delta y}\left(\frac{C(x_i, y_{j+1})}{\sqrt{x_i^2 + y_{j+1}^2}} - \frac{C(x_i, y_{j-1})}{\sqrt{x_i^2 + y_{j-1}^2}}\right) \\ & + \Gamma\left(\frac{C(x_{i+1}, y_j) - 2C(x_i, y_j) + C(x_{i-1}, y_j)}{\Delta x^2}\right. \\ & \left. + \frac{C(x_i, y_{j+1}) - 2C(x_i, y_j) + C(x_i, y_{j-1})}{\Delta y^2}\right),\end{aligned} \tag{A2}$$

where $(x_i, y_j)$ with $(i = 0, 1, \ldots 2^{N_x} - 1)$ and $(j = 0, 1, \ldots 2^{N_y} - 1)$ denotes the grid points.

With equal grid points in $x$ and $y$ $(N_x = N_y = N)$, the wavefunction $|C\rangle$ of the set of qubits is defined by having $C(x_i, y_j)$ as the $k$th element of its vector representation. The relation between $i$, $j$, and $k$ is best described using the binary notation. Let $i = (a_0, a_1, \ldots, a_{N-1})_2$ and $j = (b_0, b_1, \ldots, b_{N-1})_2$. Then, $k$ is equal to $(b_0, a_0, b_1, a_1, \ldots, b_{N-1}, a_{N-1})_2$. This is a one-to-one mapping and maps each point in a 2D grid of size $2^N \times 2^N$ to a vector with $2^{2N}$ elements. As an example, for a $16 \times 16$ grid, the 45th element of the vector representation is denoted as $C(x_i, y_j)|00101101\rangle$ with $i$ and $j$ having binary representations $j = (0110)_2 = 6$ and $i = (0011)_2 = 3$.

With the assumption of periodic boundary condition in both directions, the wavefunction transport is the same as in Eq. (3). Finding the elements of $\hat{A}$ is described in Algorithm 1 following the 2D vector representation of $|C\rangle$. For example, if $k = (11010011)_2 = 211$, this shows $i = (1101) = 13$ and $j = (1001) = 9$. Now, the nonzero elements are

$$\begin{aligned} i+1 &= (1110)_2, \quad j = (1001)_2 \\ &\implies \quad k_{i+1,j} = (11010110)_2 = 214, \end{aligned} \tag{A3a}$$

$$\begin{aligned} i-1 &= (1100)_2, \quad j = (1001)_2 \\ &\implies \quad k_{i-1,j} = (11010010)_2 = 210, \end{aligned} \tag{A3b}$$

$$\begin{aligned} i &= (1101)_2, \quad j+1 = (1010)_2 \\ &\implies \quad k_{i,j+1} = (11011001)_2 = 217, \end{aligned} \tag{A3c}$$

$$\begin{aligned} i &= (1101)_2, \quad j-1 = (1000)_2 \\ &\implies \quad k_{i,j-1} = (11010001)_2 = 209. \end{aligned} \tag{A3d}$$

Therefore, the 209th, 210th, 211th, 214th, and 217th columns of this row should be filled with the respective coefficients from the finite difference formula.

ALGORITHM 1. Constructing $\hat{A}$ in 2D.

$k \leftarrow 0$
$\hat{A} \leftarrow \text{zeros}(2^{2N}, 2^{2N})$
**while** $k < 2^{2N}$ **do**
  $(b_0 a_0 b_1 a_1 \ldots b_{N-1} a_{N-1})_2 \leftarrow k$
  $j \leftarrow (b_0 b_1 \ldots b_{N-1})_2$
  $i \leftarrow (a_0 a_1 \ldots a_{N-1})_2$
  $(b_0^+ b_1^+ \ldots b_{N-1}^+)_2 \leftarrow j+1$
  $(b_0^- b_1^- \ldots b_{N-1}^-)_2 \leftarrow j-1$
  $(a_0^+ a_1^+ \ldots a_{N-1}^+)_2 \leftarrow i+1$
  $(a_0^- a_1^- \ldots a_{N-1}^-)_2 \leftarrow i-1$
  $k_{i+1,j} \leftarrow (b_0 a_0^+ b_1 a_1^+ \ldots b_{N-1} a_{N-1}^+)_2$
  $k_{i-1,j} \leftarrow (b_0 a_0^- b_1 a_1^- \ldots b_{N-1} a_{N-1}^-)_2$
  $k_{i,j+1} \leftarrow (b_0^+ a_0 b_1^+ a_1 \ldots b_{N-1}^+ a_{N-1})_2$
  $k_{i,j-1} \leftarrow (b_0^- a_0 b_1^- a_1 \ldots b_{N-1}^- a_{N-1})_2$
  $\hat{A}[k, k_{i+1,j}] \leftarrow \frac{y_j}{2\Delta x\sqrt{x_{i+1}^2+y_j^2}} + \frac{\Gamma}{\Delta x^2}$
  $\hat{A}[k, k_{i-1,j}] \leftarrow -\frac{y_j}{2\Delta x\sqrt{x_{i-1}^2+y_j^2}} + \frac{\Gamma}{\Delta x^2}$
  $\hat{A}[k, k_{i,j+1}] \leftarrow \frac{x_i}{2\Delta y\sqrt{x_i^2+y_{j+1}^2}} + \frac{\Gamma}{\Delta y^2}$
  $\hat{A}[k, k_{i,j-1}] \leftarrow -\frac{x_i}{2\Delta y\sqrt{x_i^2+y_{j-1}^2}} + \frac{\Gamma}{\Delta y^2}$
  $\hat{A}[k, k] \leftarrow -2\Gamma\left(\frac{1}{\Delta x^2} + \frac{1}{\Delta y^2}\right)$
  $k \leftarrow k+1$
**end while**

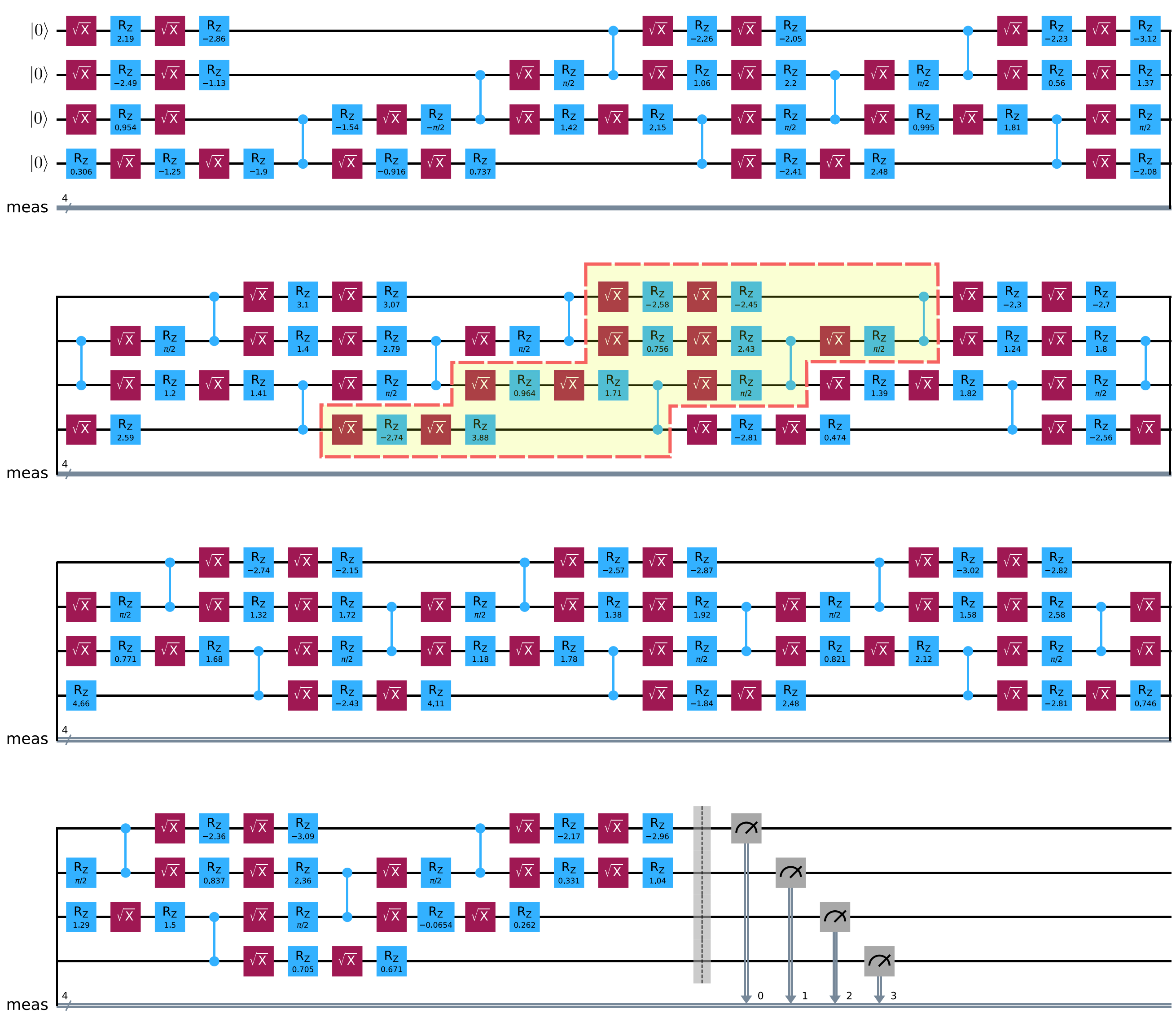

FIG. 9. Transpiled circuit of VarQTE algorithm on IBM Fez with $N = 4$ qubits and $L = 10$ layers. The highlighted area shows the structure of a single layer after transpilation into native gates.

## APPENDIX B: TRANSPILED ANSATZ FOR VARQTE AND AVQDS

In Fig. 9, the circuit for the VarQTE algorithm is shown with ten layers, transpiled into native gates. The repeating layer of gates is clearly identified. The full ansatz for the AVQDS method is shown in Fig. 10. This circuit is constructed using a predefined pool of operators. This imposed limitation on the pool reduces the number of operators used, resulting in fewer gates overall. The highlighted box in Fig. 9 indicates one layer of rotation and entanglement.

## APPENDIX C: RESOURCE ESTIMATION OF HADAMARD-TEST-TYPE CIRCUITS

The VarQTE ansatz with $N$ qubits and $L$ layers uses $N(L+1)$ variational parameters. At each time step, entries of matrix $A$ and vector $R$ defined in Eqs. (13a) and (13b) are obtained by Hadamard-test-type circuits. The dimension of matrix $A$ is $N(L+1) \times N(L+1)$. However, this is a symmetric matrix, so it needs $(N^2(L+1)^2 + N(L+1))/2$ circuits to find all its elements. The vector $R$ has $N(L+1)$ elements, and each element needs $N_H$ of circuits, where $N_H$ is the number of Pauli strings in the Hamiltonian $\hat{H}$. Thus, $(N^2(L+1)^2 + N(L+1))/2 + N_H N(L+1)$ circuits are necessary to obtain all the entries of $A$ and $R$. For $N = 4$ qubits, $L = 10$, and $N_H = 22$, this number is 2046 for every time step. Considering the final simulation time of $t = 1$ and $\Delta t = 0.002$, the total number of circuits would be $1.023 \times 10^6$.

For AVQDS, $A$ and $R$ can also be measured at each time step by the Hadamard test. This requires $N_H N_\theta + N_\theta(N_\theta + 1)/2$ different circuits, where $N_\theta$ is the number of variational parameters. In addition, $\langle \hat{H} \rangle$ and $\langle \hat{H}^2 \rangle$ must be measured to compute the McLachlan distance $D$. The overhead cost for these measurements is negligible compared to that for determining $A$ and $R$. This is the case because both expectation values can be obtained by direct measurements using only a few circuits corresponding to the commuting groups of the Pauli strings in $\langle \hat{H} \rangle$ and $\langle \hat{H}^2 \rangle$. When the ansatz adaptive process is triggered in AVQDS, an additional number of $N_P N_\theta$ circuits is required to scan the operator pool, which consists of $N_P$ Pauli strings. Further details on the measurement process can be found in Ref. [76]. For $N = 4$ qubits and $N_H = 22$, the adaptive process is activated only during the first three time steps. Subsequently, the system evolves according to a fixed ansatz, similar to VarQTE. For this fixed ansatz, the number of variational parameters is $N_\theta = 18$. The entire simulation requires a total of 988 time steps to complete, with a simulation time of $t = 1$, to achieve fidelities comparable to those of VarQTE. The total number of distinct circuits required for this process is approximately $5.6 \times 10^5$, which is almost half the number of circuits in VarQTE.

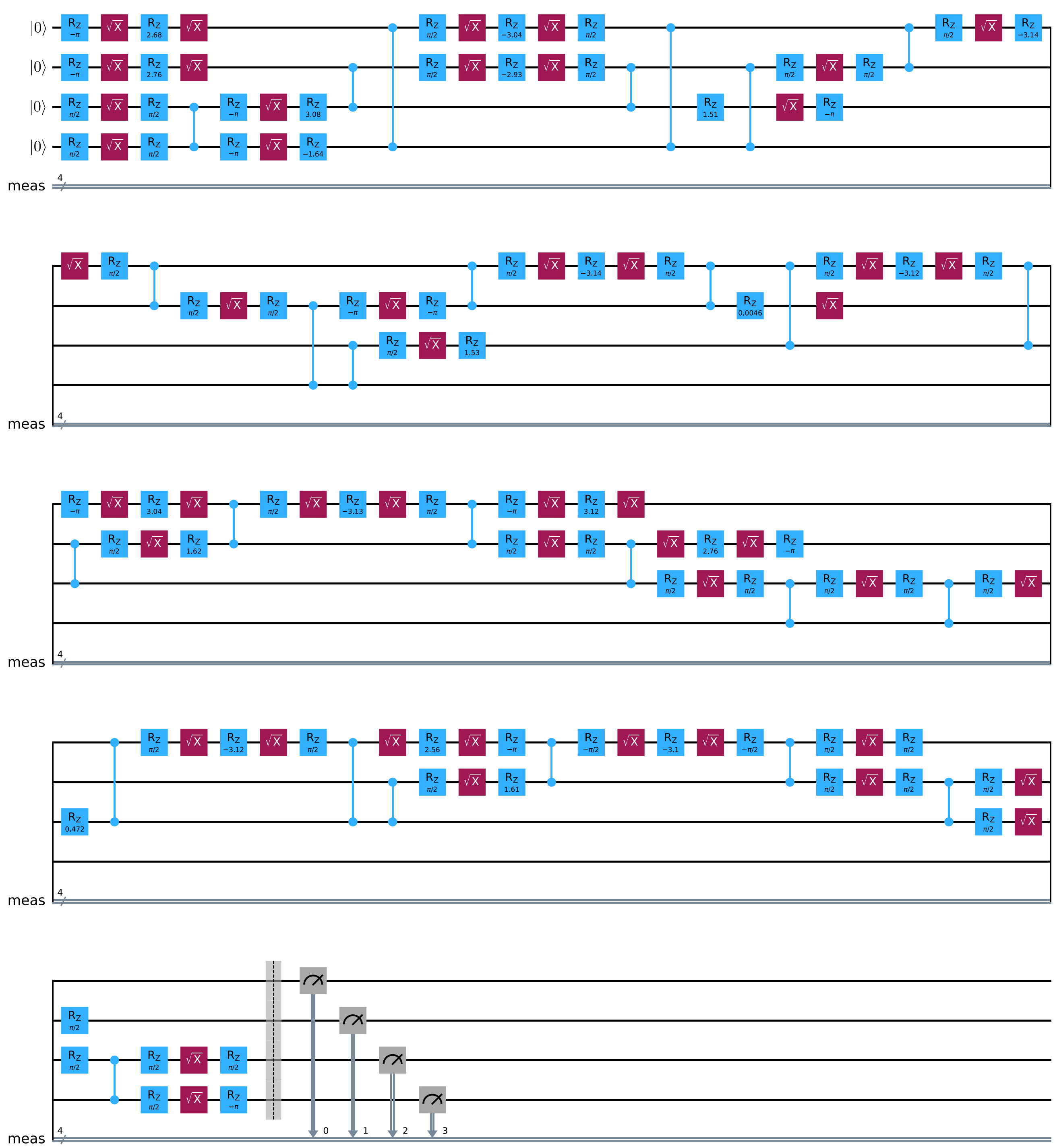

FIG. 10. Transpiled circuit of AVQDS algorithm on IBM Fez with 4 qubits.